\documentstyle[aps,prl,multicol]{revtex}
\input{epsf}
\begin{document}
\draft \title{Breakdown of universality in transitions to
  spatio-temporal chaos.}

\author{ Tomas Bohr$^1$, Martin van Hecke$^2$, Ren\'e Mikkelsen$^2$
  and Mads Ipsen$^3$} \address{$^1$Department of Physics, The
  Technical University of Denmark, DK-2800 Kgs. Lyngby, Denmark }
  \address{$^2$ Center for Chaos and Turbulence Studies, The Niels
  Bohr Institute, Blegdamsvej 17, DK-2100, Copenhagen {\O}, Denmark }
  \address{$^3$ Fritz-Haber-Institut der Max-Planck-Gesellschaft,
  Faradayweg 4-6, D-14195 Berlin, Germany \\ } \date{\today}
  \maketitle

\begin{abstract}
In this Letter we show that the transition from laminar to active
behavior in extended chaotic systems can vary from a continuous
transition in the universality class of Directed Percolation with
infinitely many absorbing states to what appears as a first order
transition. The latter occurs when {\em finite} lifetime non-chaotic
structures, called ``solitons'', dominate the dynamics.  We illustrate
this scenario in an extension of the deterministic Chat\'e--Manneville
coupled map lattice model and in a soliton including variant of the
stochastic Domany-Kinzel cellular automaton.
\end{abstract}
~\vspace{-5mm}  
~

\begin{multicols}{2}

\narrowtext 
The nature of transitions in extended deterministic dynamical systems
is not very well understood. Since few analytical methods are
available for such systems, it is tempting to map them to stochastic
models \cite{TheBook}.  In a study of the {\em deterministic} ``damped
Kuramoto--Sivashinsky equation" \cite{dKS}, Chat\'e and Manneville
\cite{CM0} introduced the notion of a universal transition to
turbulence via ``spatio-temporal intermittency" \cite{Chate,Exp}.
Pomeau gave general arguments \cite{Pomeau} for this transition to be
in the universality class of Directed Percolation (DP)
\cite{Grin1,hh}.
These arguments rested on earlier work by Grassberger \cite{Grass} and
Janssen \cite{Janss}, who conjectured that any {\em stochastic}
process with a unique absorbing state should be in the class of DP.
To check whether {\em deterministic} models could be in the DP class,
Chat\'e and Manneville introduced a very simple coupled map lattice
(CML), with the local map either performing ``laminar" or chaotic
motion.

\begin{figure} \vspace{-0.3cm}
  \epsfxsize=1.08\hsize \mbox{\hspace*{-.065 \hsize} 
\epsffile{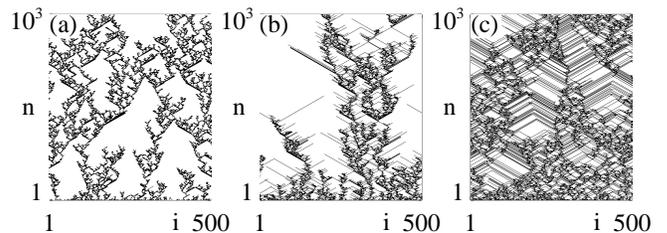}}
    \vspace{-3.8cm}
\caption[]{Spacetime plots of our coupled map lattice (\ref{CMBu})
-(\ref{CMBv}) where inactive (chaotic) sites are white (black) for
$r\!=\!3$ near criticality, illustrating the increasing role of
solitons as function of our parameter $b$: (a) $b\!=\!0.2,
\varepsilon\!=\!0.373$. (b) $b\!=\!0$, (Chat\'e-Manneville)
$\varepsilon\!=\!0.357$. (c) $b\!=\!-0.1, \varepsilon\!=\!0.352$.
}\label{fg1}
\end{figure}

Surprisingly, the critical exponents of this CML were not those of DP
\cite{CM}; in fact they appeared to vary continuously along the
critical line. Grassberger and Schreiber \cite{GS} pointed out that
the presence of long lived ``solitons" (local excitations that
propagate with unit velocity through the lattice, see Fig~\ref{fg1}b-c
and \cite{GS}) may lead to long crossover times, and conjectured that
the true asymptotic behavior of the Chat\'e-Manneville model would be
in the DP universality class.

\begin{figure}
\hspace{.3cm}
  \epsfxsize=1.27\hsize   \mbox{\hspace*{-.15 \hsize}  \epsffile{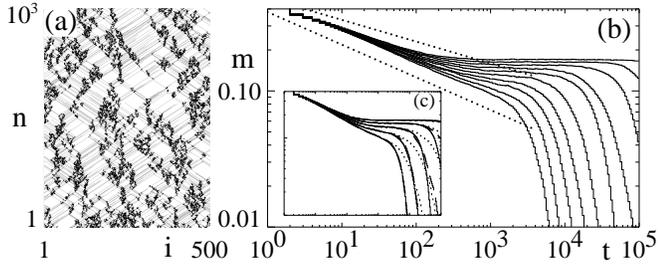} 
}
    \vspace{-4.4cm}
\caption[]{(a) Soliton dominated dynamics in the stochastic model for
$d\!=0.01$, $c\!=\!0.3$ and $p\!=\!0.55$. (b) Average activity $m$ for
$c\!=\!0.1$, $d\!=0.01$ and $p=0.612,\dots,0.621$ for an ensemble of
20 systems of size $2\times 10^4$. The dotted curves are for
comparison and have slopes $0.159$ and $0.24$ (c) Comparison of
activity for systemsizes 200 (2000 realizations, dots), 2000 (200
realizations, dashed) and 20000 (20 realizations, full curves) for
$p\!=\!0.613$, $0.615$, $0.617$, $0.619$ and $0.621$.  For smaller
systemsizes the final fall-off occuring for large $p$ seems more
gentle, but clearly the early time behavior and the value of the
plateaux activity are insensitive to finite size effects.
}\label{newf4}
\end{figure}

Our central claim is that even solitons with a {\em finite} lifetime
can completely change the nature of the transition. We base this claim
on studies of extensions of {\em{(i)}} the Chat\'e-Manneville model
\cite{CM0} and {\em{(ii)}} the Domany-Kinzel {\em stochastic} cellular
automaton \cite{DK}, where both extensions facilitate the tuning of
solitonic properties. In the regime of short soliton lifetimes the
bulk exponents of our CML are consistent with DP; the spreading
exponents indicate that the CML falls in the universality class of DP
with infinitely many absorbing states.  For both our CML and our
stochastic model the transition between active and inactive states
looses its continuous nature and {\em appears} to become first order
for large soliton lifetimes. We shall argue that this discontinuity
emerges because pairs of
\noindent 
solitons can generate new
activity upon collisions (see Fig.~\ref{fg1}c and Fig.~\ref{newf4}a)
while individual solitons do not lead to new turbulent activity.  We
study this phenomenon in detail in our stochastic cellular automaton
and show that the non-universality seen in earlier work is likely due
to the proximity of this (quasi) first order transition: our
stochastic model shows transient behaviour which can be fitted quite
convincingly to non-universal power laws.  Here we focus on the broad
picture; a detailed study will be presented elsewhere
\cite{longpaper}.

{\em Coupled Map Lattice -- } The model introduced by Chat\'e and
Manneville \cite{CM} is a 1D coupled map lattice
\begin{equation}
\label{CML}
u_i(n+1) = f(u_i(n)) +
 {\varepsilon \over 2} \Delta_fu_i(n)\, ,
\end{equation}
where $\Delta_fu_i(n)=f(u_{i-1}(n))-2f(u_i(n))+f(u_{i+1}(n))$.  When
$u \!\le \!1$, $f$ is a standard tent map of the form $f(u) = r
({\frac{1}{2}} - | u-{\frac{1}{2}}|)$ and $u$ displays chaotic
behavior, while when $u\!\ge\!1$, $f$ is simply the identity and $u$
displays laminar behavior.  States in which all sites are laminar
remain so, but chaotic sites can ``infect'' their neighbors due to the
spatial coupling. The effectiveness of this spreading of the chaos
depends on $r$ and $\varepsilon$.  Taking the density of chaotic sites
or ``activity" $m$ as an order parameter, transitions from a
``laminar" state (in which $m$ decays to zero) to a ``turbulent" state
(where $m$ reaches a finite value in an infinite system) can be
defined. To relate

\begin{figure} \vspace{-.2cm}
  \epsfxsize=1\hsize \mbox{\hspace*{-.065 \hsize} \epsffile{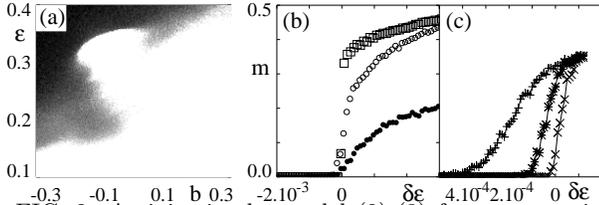} }
    \vspace{-3.4cm}
\caption[]{Activity in the model (\ref{CMBu})-(\ref{CMBv}) for a
systemsize of 2048 and ensemble of 128 runs.  (a) Activity at
$t\!=\!1000$ (white $\!=\!$ inactive). (b) Activity as function of
$\delta \varepsilon$ (distance to critical point) at time $2.10^5$ for
$b\!=\!-0.1$ (squares), $b\!=\!0$ (open circles) and $b\!=\!0.2$
(closed circles). (c) Illustration of the sharpening of the
$b\!=\!-0.1$ transition for increasing times $5.10^3$ (+), $5.10^4$
(*) and $5.10^5$ (X); note the strongly magnified scale.  }\label{fg2}
\end{figure}
\noindent
 such CML's to physical systems of weakly coupled chaotic
elements, we must interpret the map $f$ as a return map on a
Poincar\'e section; each local map should therefore be at least two
dimensional and invertible \cite{CHAOS}. We thus introduce the CML
\begin{eqnarray}
\label{CMBu}
u_i(n+1) &=& f(u_i(n)) + {\varepsilon \over 2} \Delta_fu_i(n) + v_i(n) \\
\label{CMBv}
v_i(n+1) &=& b (u_i(n+1) - u_i(n))
\end{eqnarray} 
where $f$ is the same as before and the new parameter $b$ is the
Jacobian of the local map at each site; this map is invertible for any
non-zero $b$ and becomes increasingly two-dimensional with $| b| $.
This design is analogous to the construction of the H\'enon map
\cite{Henon} from the logistic map, except that on the right hand side
$b (u_i(n+1) \!-\!  u_i(n))$ appears instead of $b u_i(n)$. This
ensures that the absorbing state fixed points $u_i(n) = u^*$ of the
old CML (\ref{CML}) are mapped to the laminar fixed point $(u_i(n),
v_i(n)) = (u^*,0)$. The model is updated synchronously and typical
dynamical states are shown in Fig.~\ref{fg1}; in what follows we will
fix the parameter $r\!=\!3$.

To get a feeling for the location and nature of the transition we show
in Fig.~\ref{fg2}a the activity after 1000 time-steps as a function of
$b$ and $\varepsilon$; the active and inactive phases can be clearly
distinguished.  In Fig.~\ref{fg2}b we show that when $b$ is varied,
there is a qualitative difference in the activity as a function of
$\varepsilon$.  The behavior for $b\!=\!0$ is consistent with a
continuous transition, whereas for $b\!=\!-0.1$ a marked steepening
occurs consistent with the emergence of a discontinuity; further
support for this comes from a study of the correlation function which
shows a finite correlation length at the transition
\cite{longpaper}. In Fig.~\ref{fg2}c we show the sharpening of this
discontinuity as function of the total integration time.

{\em CML near continuous transition -- } We focus now on the
soliton-poor regime where the transition appears to be continuous. We
have computed critical exponents using finite size scaling
\cite{TheBook,Houlrik}.  The critical line in the parameter plane
$(\varepsilon , r )$ is located by measuring the ``absorption time"
$\tau (r,\varepsilon,L)$, i.e., the average time it takes the 
 system, starting from a random initial state, to reach the

\begin{figure}
\begin{center}\label{r3}
\begin{tabular}{|l|l|l|l|} \hline 
\vspace{0cm}
 $b$     & \small $\epsilon_c$   & \small $z$   & \small
$\theta$  \\ \hline 
\small -0.25   & \small 0.16312(3)    & \small 1.58(1)& \small  \vspace{-1mm}     \\ 
\small-0.2    & \small 0.16495(2)    & \small 1.58(2)& \small 0.168(2) \vspace{-1mm}\\ 
\small-0.15   & \small 0.16205(1)    & \small 1.58(1)& \small 0.17(1)  \vspace{-1mm}\\ 
\small-0.125  & \small 0.16368(2)    & \small 1.57(1)& \small 0.20(1)  \vspace{-1mm}\\ 
\small-0.1    & \small 0.35203(1)    & \small 1.52(3)& \small 0.02(2)  \vspace{-1mm}\\ 
\small0       & \small 0.35984(3)    & \small 1.42(2)& \small 0.18(1)  \vspace{-1mm}\\ 
\small0.1     & \small 0.3393(1)     & \small 1.48(2)& \small 0.155(1) \vspace{-1mm}\\ 
\small0.125   & \small 0.34745(5)    & \small 1.53(2)& \small 0.15(1)  \vspace{-1mm}\\ 
\small0.15    & \small 0.35680(5)    & \small 1.57(1)& \small 0.159(3) \vspace{-1mm}\\ 
\small0.175   & \small 0.36545(1)    & \small 1.58(1)& \small 0.16(1)  \vspace{-1mm}\\ 
\small0.2     & \small 0.37323(1)    & \small 1.58(1)& \small 0.16(1)  \\ \hline
\small DP      & \small               & \small 1.58074(4)& \small 0.15947(3) \\ \hline
\end{tabular}
\end{center}
\vspace{-3mm}
Table 1. {\small The critical exponents $z$ and $\theta =
  \beta/\nu_\parallel$ for our CML (\ref{CMBu})-(\ref{CMBv}).  Note
  that $\varepsilon_c(b)$ is a multiple-valued function and the values
  for $b>-0.1$ correspond to the ``upper branch'' of Fig.~\ref{fg2}a.
  The values for DP (last row) are taken from \cite {Ijensen}.}
\end{figure}
\noindent absorbing state.  At the critical point
$\varepsilon=\varepsilon_c(r)$, this time diverges like $ \tau
(\varepsilon_c , L) \sim L^z$ where the usual dynamical exponent $z =
\nu_{\parallel} / \nu_{\perp}$ has been introduced. The order
parameter $m(\varepsilon,L,t)$ is the fraction of chaotic sites in the
lattice, again averaged over many different initial states. The
scaling of the order parameter, $m \sim (\varepsilon -
\varepsilon_c)^{\beta}$ for $\varepsilon \to \varepsilon_c^+$ defines
the critical exponent $\beta$.  Precisely at the critical point the
order parameter decays as $m(\varepsilon_c,t,L) \sim
L^{-\beta/\nu_{\perp}} g(t / L^z)$.  For $t \ll L^z$ this behavior
should be independent of $L$ and $m(\varepsilon_c,t,L) \sim t^{-\theta
}$ with $\theta = \beta/\nu_{\parallel}$. Table~1 shows the exponents
$z$ and $\theta$ as function of $b$ for $r\!=\!3$. These values where
obtained for systemsizes up to $L\!=\!2048$ (128 realizations)
\cite{longpaper}.  There are apparently regimes for $|b| >0.1$ where
the exponents are very close to their DP values.

\begin{figure} \vspace{-0.2cm}
\epsfxsize=1.01\hsize \mbox{\hspace*{-.07 \hsize} 
\epsffile{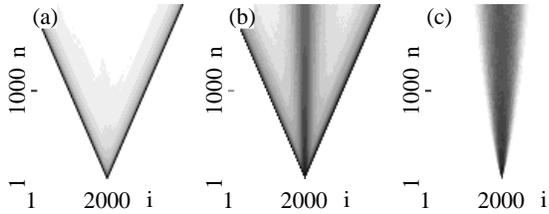} }
    \vspace{-3.cm}
\caption[]{Spreading of activity in our CML 
averaged over $10^4$ realizations for $b\!=\!-0.1$ (a), $b\!=\!-0$ (b)
and $b\!=\!0.2$ (c).}\label{fg4}
\end{figure}

{\em Spreading Properties of our CML -- } Instead of following the
decay of an initially uniformely filled system, one can also determine
critical exponents from the spreading of an initial seed of turbulence
in an otherwise laminar configuration \cite{Torre}. These so-called
dynamical exponents are defined in terms of the number of chaotic
sites $N(t)$, the survival probability $P(t)$, the mean-squared
deviation $R^2(t)$ from the origin of the turbulent activity and the
density $n(t)$ of chaotic sites within the spreading patch of
turbulence.  Thus we assume
\begin{equation}
N(t) \sim t^{\eta} \,\,\,\,\,P(t) \sim t^{-\delta} \,\,\,\,\, R^2(t) \sim 
t^{z_s}
\,\,\,\,\, n(t) \sim t^{-\theta_s}
\end{equation}
For DP one finds $\delta \!=\! \theta \!=\! \theta_s$ and $z_s \!=\!
2/z$. In the Chat\'e-Manneville model at $r\!=\!3$ it is basically
impossible to determine these exponents, since the spreading is
completely dominated by the solitons (Fig.~\ref{fg4}b). In our
generalized CML it turns out to be possible to determine the spreading
exponents (Fig.~\ref{fg4}c) in the weak soliton regime.  In the CML
the absorbing state is non-unique since any state in which all
$u$-values are above unity and all $v$-values are not too large is
absorbing. We indeed found the dynamical exponents to depend on the
configuration

\noindent of the laminar sites; the spreading exponents are
non-universal. This is consistent with recent studies of systems with
infinite numbers of absorbing states
\cite{Ijensen,JD,Mendes,Grin3,Grin2,Grass22} where it has been
revealed that these differ from classical DP precisely in the
non-universality of the spreading exponents. It has also been
conjectured, and verified numerically \cite{Mendes,Grin3}, that the
dynamical exponents satisfy the hyperscaling relation $\eta + \delta +
\theta_s =d z_s/2$ where $d$ is the spatial dimension. The spreading
exponents depend on the mean value of the field $<u>$ in the laminar
state outside of the seed.  For the example shown in Fig.~\ref{fg4}c,
the value of $<u>$ which would result from a long run with
homogeneously random initial conditions is 1.235.  Even a 2\%
variation in $<u>$ creates a 100\% variation of $\eta$ whereas $\Delta
\!\equiv\! z_s/2 -(\eta + \delta + \theta_s)$ remains less than 0.1
(for times up to 2000). Similarly, within this range $\delta$ varies
between 0 and 0.5. For other parameter values $\eta$ may range from
0.8 to 0.1; for more details, see \cite{longpaper}.

{\em First order behavior and Stochastic Model -- } The cause of the
discontinuity in the activity in the soliton rich regime is
illustrated in spacetime plots like Fig.~\ref{fg1}c: while individual
turbulent patches clearly have a finite lifetime, new activity is
created by collisions of the solitons that surround the active
patches.  To illustrate this scenario, we extend the Domany-Kinzel
cellular automaton \cite{DK} in the following way: {\em (i)} Our model
contains two species representing the chaotic sites and the solitons.
{\em (ii)} The chaotic sites behave like active sites in usual
bond-directed DP, except that when they ``die'' they can emit a left
or right-moving soliton with probability $c$.  {\em (iii)} The
solitons travel ballistically and die with a probability $d$. {\em
(iv)} Individual solitons are seen by the chaotic sites as inactive.
{\em (v)} Upon collision two solitons generate a chaotic site; this is
the only way by which solitons enhance the activity. A typical
spacetime plot of our model is shown in Fig.~\ref{newf4}a, showing the
same qualitative behavior as the CML in the soliton rich regime:
finite size clusters of activity surrounded by clouds of solitons.

{\em Mean field equations -- } The rate equations for our stochastic
model are
\begin{eqnarray}
\label{mf1}
\dot{S} &=& c C - S^2 - d S\\
\label{mf2}
\dot{C} &=& r C +  S^2 - u C^2
\end{eqnarray}
where $C$ and $S$ are the densities of chaotic sites and of solitons.
Pure DP corresponds to the equation $\dot{C} \!=\! r C \!-\! u C^2$.
These mean-field equations display both a first and second order
transition, and the nature of the transition is governed by $z
\!\equiv\! (d/c)^2 u$; the change to first order behavior occurs at
$z\!=\!1$.  If $z\!>\!1$ (short soliton life time) the solitons just
renormalize $u$ to $u(1-{\frac{1}{z}})$.  When $z<1$ the behavior for
positive $r$ (the active phase) is governed by the stable node at
$S_2^{\star}\approx a( z^{- 1/2} \!-\!1)$. When $r$ becomes negative
this fixed point remains stable and at a finite distance away from the
origin; simultaneously, the origin becomes attractive while a saddle
at $S=S_1^{\star}\! \approx\! {\frac{d r}{z-1}}$ appears close to it.
Initial conditions close to the origin will flow there, but initial
conditions above the saddle stable manifold will flow to the
node. This will go on until $(S_1^{\star},C_1^{\star})$ and
$(S_2^{\star},C_2^{\star})$ merge in a saddle-node bifurcation at $r=
r_c (z)<0$ below which the origin becomes globally attractive -
clearly a first order scenario.

{\em Activity in first order regime -- } In Fig.~\ref{newf4}b we show
the evolution of the activity $m$ for our stochastic model in the
soliton rich regime. There are two important features that can be
extracted from this data: {\em(i)} For a transient period that goes up
to time $10^3$, it is possible to find values of $p$ such that the
decay of $m$ appears to be a powerlaw with a non-DP exponent.  For the
example shown, a reasonable scaling can be obtained over 2 decades.
However, for this to be real asymptotic scaling, one should be able to
have this scaling extend to arbitrary large times; however the
activity curves for sizes 200, 2000 and 20000 precisely curve down at
the same time (Fig.~\ref{newf4}c); hence there is no hope that
increasing the systemsize extends the time interval over which
apparant scaling can be found.  {\em (ii)} For long times the activity
either decays rapidly, or first hits a plateaux. Clearly, for
increasing time the curve of $m$ as function of $p$ will make a
sharper and sharper jump, similar to what we found for the CML.  In
the plateaux regime, the qualitative dynamics is as shown in
Fig.~\ref{newf4}a. We have also checked that for $p\!=\!0.621$ the
same plateaux value is reached for initial activities in the range
from 1 to 0.1; for initial activities of $0.05$ and smaller, there is
an initial increase of the activity but the plateaux is never reached
\cite{longpaper}.

It remains to be understood whether this soliton-assisted turbulent
state can persist forever and thus whether the observed first order
transition is a true thermodynamic transition.  One can argue that
below the DP threshold (i.e., for negative $r$ in (\ref{mf2})) the
active state can be destroyed by creating a sufficiently large laminar
hole, so large that the solitons emitted from the edges cannot
penetrate through it and collide to create turbulence
\cite{Hinrichsen2}. Thus the motion of the edges of such a large
``droplet" would only be driven by the diffusion of the chaotic sites,
which for negative $r$ will cause the laminar region to expand.
Something like this is observed for states close to but below the
observed transition, where the activity remains almost constant for a
long transient time after which it rapidly decays. We have not been able to
\noindent
determine if this transient time actually diverges at the apparent
first order transition point (which would make it a bona fide phase
transition). Such an investigation is difficult since for finite
lattices the true asymptotic is always the absorbing state.

{\em Discussion} The overall picture that emerges from our models is
that the transition to spatio-temporal intermittency is strongly
influenced by coherently traveling ``solitons", which, even though
they have a finite life time, change the nature of the transition.
Recent work \cite{Elder} showing that the transition in the damped
Kuramoto-Sivashinsky equation appears first order provides additional
support for this scenario. Further support comes indirectly from the
finding \cite{Rolf} that the Chat\'e-Manneville model yields critical
behavior consistent with DP when it is driven {\em asynchronously},
which destroys the solitons. We believe that our findings also explain
the nonuniversality observed in these systems, since our stochastic
model shows transient behaviour which can be fitted convincingly to
power laws that vary strongly with the parameters \cite{longpaper}. In
general our study highlights that it is far from trivial to decide
which are the important degrees of freedom when mapping deterministic
to stochastic behavior: apparantly innoculous structures may have an
unexpected strong effect on the coarse grained dynamics and render the
``natural'', most simple stochastic models inapplicable.

It is a pleasure to thank Deepak Dhar, Martin Evans, Geoff Grinstein,
Martin Howard and Kent B{\ae}kgaard Lauritsen for valuable
discussions. MvH acknowledges financial support from the EU under
contract ERBFMBICT 972554 and support from CATS at the Niels Bohr
Institute.

\end{multicols}
\end{document}